\documentclass[runningheads]{llncs}
\usepackage[T1]{fontenc}
\usepackage{graphicx}
\graphicspath{ {images/} }
\begin{document}
\title{Yeah, I Can Play That -- Aesthetic Education in Kindergarten Using a Social Robot}
\titlerunning{Aesthetic Education in Kindergarten Using a Social Robot}
%
\author{Sinje Eggers\inst{1}\orcidID{0009-0005-7666-0827} \and
Thomas Sievers\inst{2}\orcidID{0000-0002-8675-0122}}
%
%
\institute{Hochschule für Musik und Theater Rostock, 18055 Rostock, Germany
\email{sinje.eggers@hmt-rostock.de} \and
Institute of Information Systems, University of Lübeck, 23562 Lübeck, Germany
\email{t.sievers@uni-luebeck.de}}
\maketitle              
\begin{abstract}
There have been numerous attempts to investigate the potential of social robots for education in the field of child-robot interaction (cHRI). However, empirically based models for the integration of robot technologies into aesthetic education in early childhood, especially in music education, are rare.
Our approach is based on a design-based research methodology (DBR) to investigate how a social robot can help promote the creative potential of young children, act as a supportive element, and enrich musical experiences. We therefore brought the social robot Pepper to a daycare facility for children as a companion for musical activities.
Preliminary findings suggest that the robot’s physical presence can influence attention, participation, and co-creative interaction. Given the limited data and the exploratory design approach, these results primarily serve to generate design hypotheses for subsequent iterations. In contrast to existing work, which predominantly addresses instructional or language-based learning processes, our approach focuses on co-creative musical interaction in the context of aesthetic education and audiation.

\end{abstract}
\section{Introduction}

One of the many purposes for which social robots are developed and used is education, and there have already been numerous attempts to systematically investigate their potential in this area in the field of child-robot interaction (cHRI).
Beyond that, international educational policy frameworks, particularly those of UNESCO and UNICEF, emphasize every child's right to aesthetic education, active cultural participation, and self-determined development from early childhood \cite{UNESCO2025a}. Aesthetic learning is understood as a constitutive component of a holistic education characterized by embodied, exploratory, dialogical, and participatory processes \cite{UNESCO2024}.
Despite these normative guidelines, empirically grounded models for integrating AI-based and robotic technologies into aesthetic education in early childhood -- particularly in music education -- remain scarce \cite{Ma2025,Wang2025,Martinez2023}. Previous research has primarily focused on language development or socio-emotional competencies, while musical learning processes in kindergarten have received little attention in the past.

We address this gap by investigating the potential of social robots for promoting early \textit{audiation} processes. Audiation refers to the internal listening process that enables musical thinking, understanding, and creation \cite{Gordon2013}. From an aesthetic-pedagogical perspective, learning is understood here as an open, and socially embedded activity that integrates perception, action, and reflection.
Music is conceptualized as an auditory sign system in which rhythmic and prosodic perception play a central role \cite{Nitin2023}. 
As co-creative interaction is based more on rhythmic and prosodic perception than on linguistic competence, the interaction between child and robot is considered more as an embodied developmental process than as a language-driven exchange. Established models of musical development describe successive levels of representation ranging from iconic to metric forms \cite{Gruhn2016}.
The central question is under what conditions social robots can be used in kindergartens and daycare centers without standardizing musical learning processes or replacing educational relationships. Rather, they should promote children's creative thinking and expression through situation-appropriate, dialogical interactions that playfully open up new creative spaces without supplanting the role of humans as creative forces.
Technological systems are not understood as instructive or evaluative entities, but rather as supportive media that provide creative impulses and open up spaces for aesthetic experience.

cHRI research aims to develop interactive systems that can be used in real life. 
However, implementing an inclusive educational environment with a social robot in a kindergarten requires a design that is accessible to all children and takes into account the diversity of children's behaviors, needs, and interests.
Ethics and pedagogy demand clear principles in this context: transparency of AI functions, children's participation in decisions, respect for individual creative processes, protection of privacy, promotion of social skills, and teacher accountability.
The use of social robots in a kindergarten could affect not only the privacy of the children, but also that of their educators and parents \cite{Tolksdorf2021,DiPaola}.
Beyond the legal implications, it is important to consider physical safety, as some activities in a kindergarten can be very unstructured. The psychological or mental safety of the children, inclusion, explainability, and fairness must be taken into account. Robots should not be used in kindergartens without human experts monitoring their actions and effects.

Creativity can be stimulated to a certain extent by a robot. Children who interacted with a creative robot showed higher creativity during the interaction than children who interacted with a non-creative robot \cite{Elgarf2022}. However, difficulties in arousing children's interest in a robot can easily lead to problems. A robot's inability to recognize misunderstandings is a common reason for such disruptions. Other reasons for the failure of interaction between children and a robot tutor include confusing assistance and a lack of consistency and fairness \cite{SERHOLT2018}.

This places the design of technical systems at the center of educational responsibility. Against this backdrop, a key research gap exists in the systematic investigation of the possibilities and limitations of social robotics as a supportive medium for aesthetic education in kindergarten. To date, empirically grounded design and impact models are lacking that demonstrate how social robotics can contribute to the promotion of early childhood audiation processes and what technical, didactic, and ethical prerequisites are necessary. In particular, it remains unclear what skills a social robot for education must possess to support evolutionarily structured musical developmental stages in a participatory, diversity-oriented, and child-friendly manner. The focus here is not on the substitution of pedagogical relationships, but rather on the question of how technical agents can be embedded as serving, supporting media in aesthetic learning processes. 
We are interested in the limitations, responsibilities, and specific design options for supporting these processes with social robots.

While previous research on social robots in education has primarily concentrated on language acquisition and task-oriented learning, we are specifically examining open, co-creative musical processes in early childhood. In contrast to instruction-based approaches, the robot is designed to be a responsive, non-judgmental interaction partner.
The aim of our research is to develop a learning environment, using a design-based research approach, in which a social robot is iteratively tested and analyzed with regard to its impact on early audiation processes, as well as on children's self-efficacy and participation.
Design-based research provides a suitable methodological framework for developing innovative learning environments in real-world educational contexts, refining them theoretically, and simultaneously deriving practical design recommendations.
We are investigating the extent to which social robots can promote auditory discrimination, inner hearing, and musical forms of expression such as singing, movement, and playing instruments.

\subsection{Prerequisites for Social Robotics in Early Childhood}

In order to address and support kindergarten children according to their stage of development, a selection of a suitable robot system is necessary, in addition to other prerequisites \cite{Yi2024}.
The overall goal is to give every child the opportunity to develop their musical abilities at their own pace and to exercise their right to active cultural participation. Therefore, in this work, usage of a social robot in a kindergarten is not understood as a standardizing instrument, but rather as a supporting medium for aesthetic education in line with the humanistic and educational policy guidelines of UNESCO \cite{UNESCO2025a}.
This results in important ethical requirements for our approach to robotics for education:
\begin{itemize}
    \item Principle of non-standardization: AI or robot systems must not structure musical development according to standardized performance expectations.
    \item Principle of child autonomy: Children must retain choices, room for maneuver, and opportunities for withdrawal.
    \item Principle of transparency and pedagogical control: Decisions made by the system must remain comprehensible and controllable for educators.
    \item Principle of relational sensitivity: Robots must not replace pedagogical relationships, but rather support them.
    \item Principle of data ethics: Data collection, processing and storage must be minimal, purpose-bound, and child-friendly.
\end{itemize}

These ethical criteria, together with neurobiological and music education findings, formed the analytical starting point of our design-based research process.
The main contribution of our work lies not in technical innovations, but in interaction design principles for embodied, co-creative systems.

\section{Related Work}

\subsection{Social Robotics in Early Childhood Education}
 
Social robotics is increasingly being researched as a supportive medium in the context of early childhood education, particularly in the areas of language development, socio-emotional development, and attention management. Studies show that socially adaptive robots can initiate and maintain interactions through their physical presence, multimodality, and responsiveness \cite{honghu2023}.
They achieve better learning outcomes than virtual agents and are comparable to human teachers when the tasks are simple and social interactions are required \cite{Severson}. 
Learning through social interactions is perfect for children. Kids who learn together or solve problems together retain what they have learned better and develop numerous other skills such as empathy, theory of mind, metacognition, and emotion regulation \cite{ImpactEU}. Social robots can help to improve cognitive and affective outcomes.
They are suitable as moderators in a collaborative learning process and can create a pleasant learning experience for learners \cite{Buchem23,Buchem24}. 

At the same time, it is critically discussed that many existing systems are designed to be instructional or performance-oriented and thus implicitly reproduce normative learning models \cite{Papenburg2023}. Particularly in early childhood education, there is a risk that AI-based systems will standardize developmental processes or functionally replace pedagogical relationships \cite{Rashid2023}.

The tasks of a robot in education can range from simple conversations on topics chosen by the children themselves, to language training and the development of social skills, to supporting lessons by repeating and reinforcing learning content \cite{Belpaeme2013,CHENG2018,SieversAAAI}. All these approaches share the insight that learning can be promoted in many different ways, as learners can participate in a variety of activities that support learning. 
A major challenge in using social robots in education is to combine the robot-centered perspective, i.e., the technical capabilities of the robots, with the child-centered perspective, which represents how the child or children can benefit from the robot and how the robot should behave in order to best support them in achieving their interaction goals \cite{Rudenko}. 
The nature of the social relationship between humans and robots is particularly important for language acquisition as a social endeavor \cite{Rohlfing2022}.

According to developmental psychology, problem-solving skills, cognitive flexibility, and metacognition belong to the domain of executive functions. Executive functions refer to a set of adaptive, goal-directed, top-down mental processes that are necessary when one needs to concentrate and be attentive and an automatic response is not sufficient. Robots can be used in education to improve planning and control skills for complex tasks in early childhood and promote the development of executive functions \cite{DILIETO2017,Charisi2020}.
Charisi et al. demonstrated that a robot's reliability shapes a supportive relationship between children and the robot, while the robot's expressiveness influences children's perceptions of the robot's supportive abilities and companionship. Children who interacted with the reliable robot performed better on tasks, while children who interacted with the unreliable robot exhibited more task-related social interactions \cite{Charisi2021}.

Angeli et al. reported an improvement in the computational thinking of boys and girls with statistically significant learning progress when learning with a robot \cite{ANGELI2020}.
When multiple children interact with the robot, their perceptions of it remain individual, although successful joint task completion appears to strengthen the children's perception of the robot in terms of friendship and reliability \cite{Escobar-Planas2022}.
Ahmad et al. examined the question which characteristics of a robot can lead to long-term social engagement being maintained during cHRI.\cite{Ahmad2017}.
Personalizing interaction is difficult in real-world learning environments. Studies have demonstrated the feasibility of using social robots in real-world educational settings, but have also highlighted how difficult it is to achieve long-term, highly autonomous interaction between robots and children \cite{WOO2021}.

Previous research has mainly focused on cognitive or linguistic outcomes. The use of social robotics in the context of aesthetic and musical education -- especially with regard to open, co-creative learning processes -- has so far only been investigated to a limited extent.

\subsection{Aesthetic and Musical Education in Early Childhood}

Research on aesthetic education in early childhood describes musical learning as an embodied, process-oriented, and socially embedded activity that is not primarily oriented toward formal performance goals but rather toward perception, exploration, and expression in the context of early childhood musical development \cite{Puehringer2019,Gembris2024}.
Music education approaches emphasize pre-symbolic forms of learning such as singing, movement, and improvisational play as key pathways to musical development and cultural participation in early childhood.

A central theoretical concept in this regard is \textit{audiation}, which is understood as the ability to internally hear, remember, and anticipate music \cite{Gordon2012}. Audiation is considered a fundamental prerequisite for musical thinking and develops before symbolic notation. Early childhood musical education processes therefore focus less on reproduction and more on developing internal auditory representations and dialogical musical interaction. These theoretical foundations are well established. However, there are currently few empirically grounded models that investigate how digital systems or social robots can support such pre-symbolic, audiation-related processes without undermining their openness and developmental logic.
Although audiation forms the theoretical basis of our approach, it was not directly measured during this exploratory phase. Instead, observable indicators such as rhythmic imitation, vocal variation, and participation in musical interaction were used as initial points of reference.

\subsection{Human-Robot Interaction (HRI) and Creativity}

Research on human-robot interaction emphasizes that co-creative processes are successful when robots do not act as dominant actors, but rather as responsive, adaptive interaction partners. Creativity is understood not as an output, but as an emergent process arising from situational interaction \cite{Burmester2019}.
This perspective is particularly relevant in early childhood contexts, as creative processes are closely linked to autonomy, exploration, and social negotiation \cite{Hedderich2019}. Multimodality, openness in interaction, and situational adaptation to children's signals are particularly important here.
However, it remains unclear what specific capabilities and design principles an educational robot must possess in order to support co-creative musical processes in the sense of aesthetic education without controlling or evaluating them.

\subsection{Ethical Requirements for Social Robots in Early Childhood Education}

International guidelines, in particular the UNESCO recommendations on the ethics of artificial intelligence, formulate key principles for the field of education, such as non-nomination, protection of children's autonomy, transparency, and pedagogical control \cite{UNESCO2025b}. These principles are of particular importance in the early childhood context, as children are considered a particularly vulnerable group. While the ethical guidelines are well-developed, empirical studies that systematically translate these principles into concrete design and interaction models for social robotics in kindergarten are still lacking \cite{Tolksdorf2021}.
There is also a lack of research-based design models that integrate ethical requirements, findings from music education, and technological possibilities.

\section{Methods}

\subsection{Design \& Research Approach}

Our approach follows a design-based research (DBR) methodology with the aim of developing and empirically investigating innovative learning environments in real-world educational contexts, while simultaneously generating theoretical insights. It is structured as an iterative, three-phase design process:
\begin{enumerate}
    \item \textbf{Iteration 1} (exploratory pilot phase): This paper reports on Iteration 1, which focuses on feasibility, initial interaction patterns, and the identification of preliminary design principles for the interaction between children and robots in early music education. These findings are based on exploratory observations and will be used to further refine the design.
    \item \textbf{Iteration 2} (Design Refinement Phase): Building on the results of Iteration 1, the second cycle focuses on improving the robot’s multimodal responsiveness, timing of interactions, and expressiveness. In addition, a structured observation and encoding framework will be developed to enable a systematic analysis of interaction quality and engagement.
    \item \textbf{Iteration 3} (Evaluation Phase): Iteration 3 involves a comprehensive evaluation using video-based analysis, qualitative encoding, and triangulation with feedback from educators. The goal of this phase is to validate and refine the proposed design principles and to examine their impact on aesthetic learning processes.
\end{enumerate}


Based on the theoretical frameworks of aesthetic pedagogy, music education (particularly the concept of audiation), and ethical principles for AI in education (e.g., non-standardization, child autonomy, transparency, and data minimization), key design decisions are made, including the use of non-instructive interaction patterns, multimodal interaction (movement, rhythm, sound), and the deliberate positioning of the robot as a supportive, non-judgmental companion. 
In the DBR process, empirical observations and theoretical assumptions are continuously linked: observations from real-world interaction inform design modifications, which are then iteratively tested and refined. 
The goal is not only to improve the specific learning environment but also to derive transferable design principles for the interaction of social robots in aesthetic education during early childhood. The DBR approach was chosen because it is particularly well-suited for the development, testing, and theory-guided refinement of innovative educational offerings in real-world pedagogical contexts. This format combines empirical research with design-oriented development and aims to generate both practice-relevant results and theoretically grounded models. The robot does not function as a teaching or evaluation authority, but rather as an interactive catalyst that initiates, varies, and contextually accompanies musical experiences.

This approach addresses the challenge of enabling collaborative creative processes without allowing robotic systems to dominate the interaction. Design decisions explicitly prioritize the child's initiative, educator guidance, and the open-ended nature of musical exploration. Modularity is considered a central design strategy, not a technical feature. 
The interaction design integrates verbal, gestural, auditory, and physical channels to foster the nonverbal and embodied forms of musical expression typical of early childhood learning, which could reduce reliance on language skills and promote inclusive participation.
Ethical principles such as child autonomy, freedom of choice, and data privacy are directly integrated into the interaction design.
Transparency is achieved through interaction patterns that enable children and educators to develop an experience-based understanding of the social robot's capabilities and limitations, thereby fostering a reflective engagement with the technology.
The design outcome consists of modular interaction elements and musical impulse structures that can be reused and adapted in various educational settings. Examples include adaptive musical prompts, rhythm-based question-and-answer structures, and interaction patterns with alternating dialogue. Within the DBR process, the modules are iteratively refined and evaluated through participatory feedback loops that systematically incorporate children's perspectives. This approach provides transferable design guidelines and practical recommendations for educators working with social robots in early childhood music education.
Key design principles are:
\begin{itemize}
    \item Audience orientation: Focus on singing, movement, rhythmic and tonal
variation before symbolic fixation.
    \item Adaptivity without evaluation: Adaptation to children's reactions without
performance-related feedback.
    \item Aesthetic openness: Ambiguous musical offerings instead of goal-oriented
tasks.
    \item Participation: Children determine the duration, intensity, and form of the interaction.
    \item Ethical-by-design: Ethical criteria are an integral part of the technical and
didactic design.
\end{itemize}

Our work to date comprises Iteration 1 and focuses exclusively on the exploratory testing of an initial interaction design without systematic data analysis. No controlled variables were altered; instead, we gathered initial observations regarding the dynamics of interaction.
The findings from this phase pertain in particular to: 
\begin{itemize}
    \item the role of the robot as a companion versus an instructor,
    \item the importance of nonverbal interaction,
    \item technical limitations on perception and responsiveness
\end{itemize}

\subsection{Participants and Setting}

\begin{figure}
\centering
\includegraphics[width=0.47\textwidth]{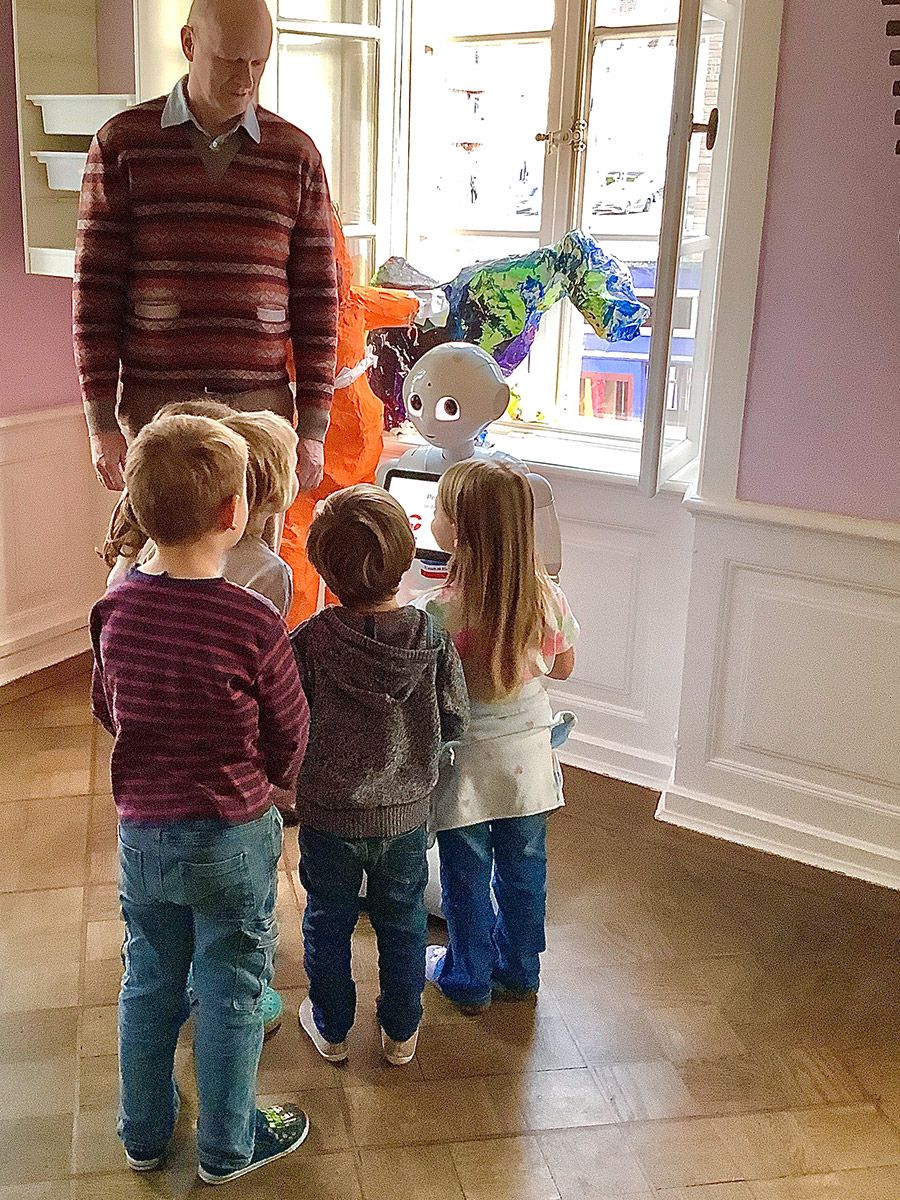}
\caption{Pepper robot interacting with children in a daycare facility.}
\label{interaction}
\end{figure}

Our observations to date took place in a daycare facility for children in the city of Lübeck with the social robot Pepper, which served as a companion for musical activities.
The observations were conducted during an initial exploratory phase with children aged 3 to 6 years.
The test phase has so far consisted of two on-site appointments with several groups of children participating in succession, with each group consisting of approximately 6 to 10 children.
The parents of the participating daycare facility children were informed in advance about the objectives, procedure, and framework of the research project and consented to their children's participation. Fig.~\ref{interaction} shows Pepper with children in the daycare facility listening to the robot.

The humanoid social robot Pepper used in our scenarios was developed by Aldebaran and first released in 2015 \cite{Pepper}. The robot is 120 centimeters tall and optimized for HRI. It is able to engage with people through conversation, gestures and its touch screen. The robot features an open and fully programmable platform so that developers can program their own applications using software development kits (SDKs) for programming languages like C++, Python or Java respectively Kotlin \cite{PepperSDK2023}.

When considering how interaction between the robot and children might work, we evaluated the use of a Large Language Model (LLM) for language generation against the creation of a scripted sequence plan with predefined dialogues.
LLMs, such as those from OpenAI or the European Mistral AI, have been widely used for some time to generate comprehensive language capabilities for robots via their application programming interface (API). LLMs also have the potential to improve teaching, learning, and administrative processes in the education sector \cite{Melzer2025}.
Vision-Language Models (VLMs) supplement these capabilities with a sense of sight, enabling the robot to interpret and describe visual impressions in natural language, thereby embedding it more firmly in the world of its human conversation partner \cite{ghosh2024}. Such multimodal models can make children's actions more understandable to the robot by providing additional visual information without the children having to name them.

\begin{figure}
\centering
\includegraphics[width=0.62\textwidth]{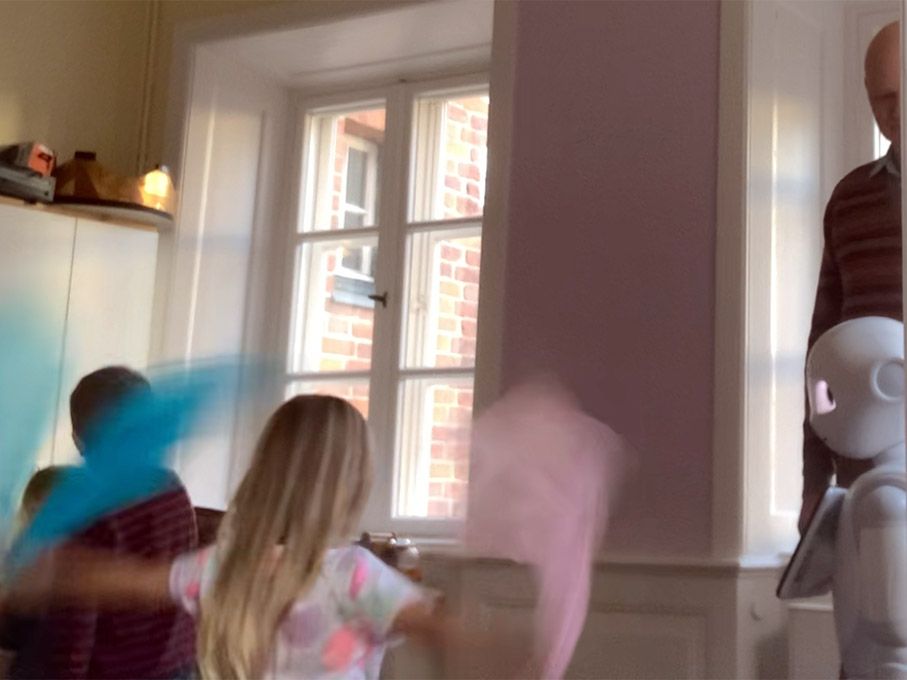}
\caption{Children move to music played by Pepper.}
\label{dancing}
\end{figure}

Nevertheless, for our first attempts at interactions with the children, we decided on a fixed procedure in which the individual steps, including the robot's contributions to the conversation, were predetermined. We adjusted the robot's voice pitch and speaking speed slightly for the target audience to improve comprehensibility.
For a musical exercise with claves, for example, Pepper used audio to play different rhythms that the children then had to reproduce themselves. These rhythms had different levels of difficulty, and the children could decide whether they wanted to repeat the exercise or move on to the next one by responding to a corresponding question from the robot.
Pepper explained the individual steps of the exercise and encouraged the children based on their progress.
Another time, Pepper played different pieces of music to which the children were supposed to move around the room, sometimes using colorful scarves or balloons as well, as can be seen in Fig.~\ref{dancing}.

\begin{figure}
\centering
\includegraphics[width=0.47\textwidth]{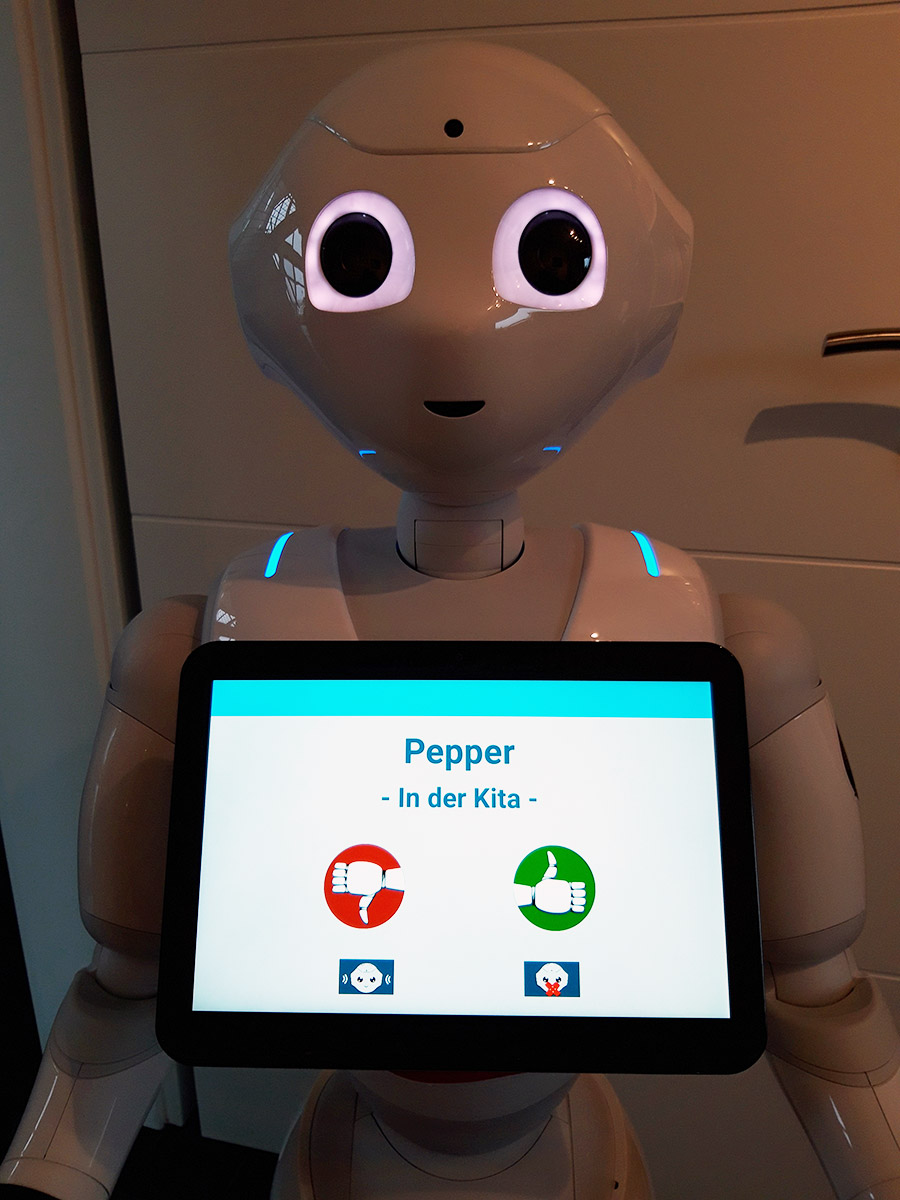}
\caption{Pepper robot with a yes button (green) and a no button (red) on its tablet.}
\label{yes-no}
\end{figure}

Since automatic speech recognition engines are often unable to reliably recognize children's speech -- and this is also the case with Pepper's speech-to-text processing -- we have decided to add an additional nonverbal interaction option \cite{Kennedy2017}.
We used the robot's tablet to display a \textit{yes} and \textit{no} button after each question, which the children could simply press to send their answer to the robot (see Fig.~\ref{yes-no}).
The questions were formulated accordingly so that a yes or no answer was sufficient. This also avoided the problem of poor comprehension due to multiple answers from a group.

\subsection{Framework for Data Collection and Monitoring}

This work is based on exploratory, non-standardized, yet theory-guided observations conducted during the first two sessions and aligned with the objectives of the first DBR iteration. No formal encoding scheme was used during this phase. The researchers documented their observations during and immediately after the sessions using structured field notes. 
The observations focused on the following indicators, which serve as a preliminary basis for identifying relevant interaction patterns and are to be further refined in future DBR cycles:

\begin{itemize}
    \item Duration of attention,
    \item the period of time during which a child actively focuses their attention on the robot or the musical activity,
    \item the frequency of reciprocal interaction between the child and the robot,
    \item number of interactions initiated by children (such as through spontaneous musical contributions, addressing the robot, or self-initiated movements),
    \item forms of multimodal participation (combination of vocal, motor, and instrumental forms of expression within an interaction),
    \item co-creative sequences in which the child and robot engage in musical interaction that is mutually responsive (e.g., imitation, variation, rhythmic dialogue)
\end{itemize}


\section{Preliminary Findings and Discussion}

This work contributes to basic research as well as to pedagogical practice in early childhood education.
Our concept is not intended as a finished product, but rather as a preliminary
design artifact that will be tested and further developed in educational practice.
Based on evaluation results, the design will be iteratively revised. 
The redesign phase should focus on theory-driven development and modeling that incorporates both empirical findings and theoretical considerations from the fields of music education, neurobiology, technology ethics, and aesthetic education.
The goal of the initial early phase was to develop a theoretically sound and practically relevant model that describes the following:
\begin{itemize}
    \item Under what conditions can a social robot support early audiation processes?
    \item Which design principles have proven effective?
    \item Where are the limits of technological support for aesthetic education?
\end{itemize}

The preliminary results are based on the first exploratory observation phase and do not yet allow for generalizable conclusions. However, they provide indications of key interaction patterns between children and the social robot Pepper. Initial observations showed that the robot's physical presence supported the children's focus of attention and cooperative participation.
These observations of interactions comprised indicators of social engagement, such as shared attention, conversation skills, and proxemics. Indicators of musical participation, such as vocal, motor, and instrumental activity, were also included. At the same time, the interaction patterns among peers, the children's cooperative behavior with the social robot, and the social behavior within the group were observed.

Multimodal signals on the part of the robot, such as gestures, facial expressions, and speech, were rudimentary and need to be further differentiated in order to contribute to the stabilization of joint musical activities. 
Another robot model with more expressive facial expressions, such as the Navel from navel robotics GmbH, might be more suitable for this purpose \cite{navel}.
Particularly strong participation from the children was observed when the robot acted not as an instructor but as a socially responsive companion. 
This was particularly the case when the robot played songs and when the children engaged in free play and movement with colorful scarves.
In this situation, the children increasingly made improvisational contributions, such as vocal variations, rhythmic movements, or instrumental experimentation. 
These behaviors must be recognized more quickly and clearly by the social robot via its perception modules. Only then is the prerequisite created for a co-creative musical process, which alternates with phases of shared attention between child, robot, peer group, and educator.

\section{Conclusion and Future Work}

We have examined the question of how robot technologies can be incorporated into early childhood aesthetic education, particularly music education, and how a robot can help to promote children's creative potential, act as a supporting element, and enrich musical experiences.
Preliminary observations can be described based on micro-analytical indicators such as attention span, frequency of turn-taking sequences, and self-initiated interactions among the children, even though no systematic quantification was performed in the first iteration of our approach. These exploratory observations suggest that the robot’s physical presence may have fostered the children’s attention, cooperative participation, and joint musical improvisation. The impression was that the children participated more actively when it came to joint musical improvisation and the robot acted as a social companion rather than an instructor.

In addition to improved immediate responsiveness to the children's behavior, another skill that the robot should retain is the ability to move in order to dance together with the children.
Improved and more flexible cognitive abilities with regard to language, visual inputs, and also memory would be interesting features for expanding and enhancing the possibilities for interaction. In this regard, improving the robot's capacity for empathetic behavior would also be useful.
In the next iteration (Iteration 2), the following specific enhancements will be implemented to systematically test our hypotheses:
\begin{itemize}
    \item Development of a structured observation and encoding scheme,
    \item improvement of multimedia perception (particularly movement and rhythm),
    \item adjustment of the robot’s interaction timing and responsiveness,
    \item comparison of different interaction modes (more open vs. more structured)
\end{itemize}

%
%
%
\bibliographystyle{splncs04}
\bibliography{references}

\end{document}